\documentclass{blois}

\def\Journal#1#2#3#4{{#1} {\bf #2}, #3 (#4)}

\def\PLB{{\em Phys. Lett.}  B}
\def\PRL{\em Phys. Rev. Lett.}
\def\PRD{{\em Phys. Rev.} D}

\def\be{\begin{equation}}
\def\ee{\end{equation}}
\def\bea{\begin{eqnarray}}
\def\eea{\end{eqnarray}}

\newcommand{\Photo}{}

\usepackage{amsmath}

\begin{document}
\vspace*{4cm}
\title{Light hyperon physics at the BESIII experiment}

\author{ Varvara Batozskaya }

\address{Institute of High Energy Physics, Beijing 100049, People's Republic of China\\
National Centre for Nuclear Research,  Pasteura 7, 02-093 Warsaw, Poland}

\maketitle\abstracts{
The BESIII experiment at the electron-positron collider BEPCII in Beijing (China) is successfully operating since 2008 and has collected large data samples in the tau-mass region, including the world's largest data samples at the $J/\psi$ and $\psi'$ resonances. The recent observations of hyperon polarizations at BESIII opens a new window for testing CP violation, as it allows for simultaneous production and detection of hyperon and anti-hyperon pair two-body weak decays. The CP-symmetry tests can be performed in processes like e.g. $J/\psi\to\Lambda\bar{\Lambda}$, $J/\psi,\psi'\to\Sigma^+\bar{\Sigma}^-$ and $J/\psi\to\Xi^-\bar{\Xi}^+$. For the $\Xi^-\to\Lambda\pi^-$ decay it is possible to perform three independent CP tests and determine the strong phase and weak phase difference. }

\section{Introduction}

One of the unresolved questions of fundamental physics is why there exists a strong abundance of matter over anti-matter in the Universe. According to the current paradigm there existed an equal amount of anti-matter at the Big Bang. The matter-antimatter asymmetry is therefore expected to have arisen via a physical mechanism, called baryogenesis~\cite{sak}, which is required the violation of charge conjugation (C) and charge conjugation combined with parity (CP) in the processes. 
The CP symmetry allowed by the Standard Model (SM) is not sufficient to account for the observed discrepancy between
matter and anti-matter. Thus, if an enhanced CP violation were to be measured this would indicate new physics and provide a clue to what happened to the missing anti-matter.

If hyperons are polarized, the direct tests on CP symmetry can be conducted by simultaneously measuring the angular distributions of the hyperon and anti-hyperon decay products. Since any CP-violating effect is small, high precision is required. It is therefore a necessity that large data samples are available. Precise CP tests on hyperon-antihyperon pairs can be performed in the processes $e^+e^-\to J/\psi,\psi'\to B\bar{B}$. The BESIII collaboration~\cite{abl0} has collected the world's largest data sample directly from electron-positron annihilation, and allows for several stringent precision tests on CP symmetry. So far the released analyses are based on $1.3\times10^9$ $J/\psi$ and $4.5\times10^7$ $\psi'$ events but there is more data are available, $10^{10} J/\psi$ and $3\times10^9$ $\psi'$. 

\section{Hyperon decays}\label{subsec:prod}

The main decay modes of the ground-state hyperons are the weak $\Delta S=1$ transitions into a baryon and a pseudoscalar meson. Historically they provided the crucial information for establishing the pattern of parity violation in weak decays~\cite{lee1}. Nowadays they are used in searches of CP-symmetry violation signals in the baryon sector and to determine spin polarization in hadronic reactions involving hyperons.

For a weak decay of a spin-1/2 mother ($B$) baryon to a spin-1/2 daughter ($b$) baryon and a pion, like $\Lambda\to p\pi^-$ or $\Xi^-\to\Lambda\pi^-$, the parity-even (parity-odd) amplitude leads to the final state in the $p$-wave ($s$-wave). The two amplitudes denoted $P$ and $S$, respectively, can be parametrized using two independent decay parameters~\cite{lee2}:
\begin{equation}
 \alpha_D=\frac{2\mathrm{Re}(S\ast P)}{|S|^2+|P|^2}\qquad\mathrm{and}\qquad\beta_D=\frac{2\mathrm{Im}(S\ast P)}{|S|^2+|P|^2}, 
\label{eq:alpphi} 
\end{equation}
where $|S|^2+|P|^2$ is the normalisation of amplitudes. The parameters provide the real and imaginary part of the interference term between the amplitudes. The experimentally motivated parameter is $\phi_D$, $\phi_D\in[-\pi,\pi]$, which is related to the rotation of the spin vector between mother and daughter baryons. For $\Xi^-\to\Lambda\pi^-$ decay with polarized cascade the $\phi_D$ parameter can be determined using the subsequent $\Lambda\to p\pi^-$ decay which acts as a polarimeter. The relation between $\beta_D$ and $\phi_D$ parameters is $\beta_D=\sqrt{1-\alpha^2_D}\sin\phi_D$. The decay parameter $\alpha_D$, $\alpha_D\in[-1,1]$, can be determined from the angular distribution asymmetry of the $b$ baryon in the $B$ baryon rest frame. The distribution is given as
\begin{equation}
\frac{1}{\Gamma}\frac{\mathrm{d}\Gamma}{\mathrm{d}\Omega}=\frac{1}{4\pi}(1+\alpha_D{\bf P}_B\cdot{\bf\hat{n}}),
\label{eq:dist}
\end{equation}
where ${\bf P}_B$ is the $B$ baryon polarization vector and ${\bf\hat{n}}$ is the direction of the $b$ baryon momentum in the $B$ baryon rest frame. In the CP-conserving limit the hyperon-antihyperon average values can be defined as $\langle\alpha_D\rangle=(\alpha_D-\bar{\alpha}_D)/2$ and $\langle\phi_D\rangle=(\phi_D-\bar{\phi}_D)/2$. Experimentally, two independent CP-violation tests based on comparison of the decay parameters in the hyperon and anti-hyperon processes are possible~\cite{pai}
\begin{equation}
 A_{\mathrm{CP}}^D=\frac{\alpha_D+\bar{\alpha}_D}{\alpha_D-\bar{\alpha}_D}\qquad\mathrm{and}\qquad\Phi_{\mathrm{CP}}^D=\frac{\phi_D+\bar{\phi}_D}{2}.
 \label{eq:APhi1}
\end{equation}

The two-body hyperon decay can be described using decay matrices $a_{\mu\nu}^D$ representing the transformations of the spin operators (Pauli matrices) $\sigma_{\mu}^B$ and $\sigma_{\nu}^b$ defined in the $B$ and $b$ baryon rest frames, respectively~\cite{per}:
\begin{equation}
\sigma_{\mu}^B\to\sum_{\nu=0}^3 a_{\mu\nu}^D\sigma_{\nu}^b.
\label{eq:amunu}
\end{equation}
The elements of such $4\times4$ matrices are parameterized in terms of the decay parameters $\alpha_D$ and $\phi_D$ and depend on the helicity angles.

\section{Production of baryon-antibaryon pairs and joint angular distibutions}

Thanks to the relatively large branching fractions~\cite{zyl}, and low hadronic background, the $e^+e^-\to J/\psi,\psi'\to B\bar{B}$ processes are well suited for determination of hyperon decay properties and CP-violation tests. Two analysis methods are possible: exclusive where the both decay chains of baryon and anti-baryon are fully reconstructed and inclusive where only the decay chain of the baryon or anti-baryon is reconstructed. The importance of all single-step weak decays, e.g. the $\Lambda\to p\pi^-$, is that the $\Lambda$ and $\bar{\Lambda}$ are produced with a transverse polarization. The polarization and the spin correlations allows for a simultaneous determination of $\alpha$ and $\bar{\alpha}$ decay parameters~\cite{fal}.

To describe the baryon-antibaryon pair production in electron-positron annihilation including the two-body sequential decay processes a modular approach~\cite{per} can be used. The general expression for the joint density matrix of the $B\bar{B}$ pair is:
\begin{equation}
 \rho_{B\bar{B}}=\sum_{\mu,\nu=0}^3 C_{\mu\nu}\sigma_{\mu}^B\otimes\sigma_{\nu}^{\bar{B}},
\label{eq:rhoBB} 
\end{equation}
where a set of four Pauli matrices $\sigma_{\mu}^B(\sigma_{\nu}^{\bar{B}})$ in the $B(\bar{B})$ rest frame is used and $C_{\mu\nu}$ is $4\times4$ real matrix representing polarizations and spin correlations. It describes the spin configuration of the entangled hyperon-antihyperon pair in their respective helicity frames. The coefficients $C_{\mu\nu}^{1/2}$ depend on the angle $\theta$ between the positron and baryon $B$. The structure of the $C_{\mu\nu}^{1/2}$ $4\times4$ matrix can be represented by polarization vector 
\begin{equation}
 P_y(\theta)=\frac{\sqrt{1-\alpha^2_{\psi}}\sin\theta\cos\theta}{1+\alpha_{\psi}\cos^2\theta}\sin(\Delta\Phi)
\label{eq:Py} 
\end{equation}
and spin correlations $C_{ij}^{1/2}(\theta)$. The $\alpha_{\psi}$ and $\Delta\Phi$ are two real parameters describing the angular distributions of the baryon-antibaryon pair production. The complete joint angular distribution of $J/\psi\to B\bar{B}$ with a single-step decay of hyperon and anti-hyperon is
\begin{equation}
\mathcal{W}(\boldsymbol\xi;\boldsymbol\omega)=\sum_{\mu,\nu=0}^3 C_{\mu\nu}^{1/2}a_{\mu0}^Da_{\nu0}^{\bar{D}}.
\label{eq:Wss}
\end{equation}
The vector $\boldsymbol\xi$ represents a complete set of helicity angles. The complete global parameter vector $\boldsymbol\omega$ has four dimensions: $\boldsymbol\omega=(\alpha_{\psi},\Delta\Phi,\alpha_D,\bar{\alpha}_D)$. 

The production and the two-step decays in the $e^+e^-\to J/\psi\to\Xi^-\bar{\Xi}^+$ are described by a nine-dimensional vector $\boldsymbol\xi$ of the helicity angles. The structure of the nine-dimensional angular distribution is determined by eight global parameters $\boldsymbol\omega_{\Xi}=(\alpha_{\psi},\Delta\Phi,\alpha_{\Xi},\phi_{\Xi},\alpha_{\Lambda},\bar{\alpha}_{\Xi},\bar{\phi}_{\Xi},\bar{\alpha}_{\Lambda})$~\cite{per}.

\section{Experimental measurements}
\subsection{$e^+e^-\to J/\psi\to\Lambda\bar{\Lambda}$}

The first observation of the polarization in $e^+e^-\to J/\psi\to\Lambda\bar{\Lambda}$ at BESIII has been reported in 2018~\cite{abl1}. The final data samples include 420,593 and 47,009 candidates with an estimated background events of $399\pm20$ and $66\pm8$ for the $p\pi^-\bar{p}\pi^+$ and $p\pi^-\bar{n}\pi^0$ final states, respectively. A clear signal polarization dependent on the $\Lambda$ direction is observed for $\Lambda$ and $\bar{\Lambda}$. The phase between helicity flip and helicity conserving transitions is determined to be $\Delta\Phi=(42.4\pm0.6\pm0.5)^{\circ}$. This value of the phase corresponds to the transverse polarization $P_y$ (Eq.~\ref{eq:Py}) reaching maximum of 25$\%$. The obtained value of $\langle\alpha_{\Lambda}\rangle$ is $0.754\pm0.003\pm0.002$ deviating by $17\%$ from the world average established forty years ago for the $\alpha_{\Lambda}=0.642\pm0.013$~\cite{tana}. The CLAS experiment~\cite{ire} has re-analyzed spin data on $\gamma p\to\Lambda K^+$ and found the value of $\alpha_{\Lambda}=0.721\pm0.006\pm0.005$ which is still inconsistent with the BESIII result that needs to be understood.

\subsection{$e^+e^-\to J/\psi,\psi'\to\Sigma^+\bar{\Sigma}^-$}

The value of the decay parameter $\alpha_{\Sigma}$ in the process $\Sigma^+\to p\pi^0$ prior to the BESIII measurement performed in 2020~\cite{abl2} was based on the $\pi^+p\to\Sigma^+ K^+$ experiments fifty years ago~\cite{har,bel,lip} while $\bar{\alpha}_{\Sigma}$ has not been measured. The process $e^+e^-\to J/\psi,\psi'\to\Sigma^+\bar{\Sigma}^-$ is interesting in the context of revealing quantum entangled spin correlations since the large $|\alpha_{\Sigma}|$ value enhances sensitivity. The value of $\alpha_{J/\psi}$ parameter is determined to be negative and $\alpha_{\psi'}$ is positive. The relative phases between the helicity amplitudes in the $J/\psi$ and $\psi'$ decays have opposite signs and different magnitudes. The CP-odd observable $A_{\mathrm{CP}}^{\Sigma}=-0.004\pm0.037\pm0.010$ is extracted for the first time and is consistent with the SM prediction~\cite{tan}, $A_{\mathrm{CP}}^{\Sigma}\sim3.6\times10^{-6}$.

\subsection{$e^+e^-\to J/\psi\to\Xi^-\bar{\Xi}^+$}

The analysis results of the $J/\psi\to\Xi^-\bar{\Xi}^+\to(\Lambda\to p\pi^-)\pi^-(\bar{\Lambda}\to\bar{p}\pi^+)\pi^+$ process has been presented in the recently submitted manuscript~\cite{abl3}. After applying all selection criteria, 73,244 $\Xi^-\bar{\Xi}^+$ candidates remain in the sample with the remaining background of $187\pm16$ events. 

The set of weak decay parameters $\Xi^-\to(\Lambda\to p\pi^-)\pi^-$ and the production related parameters $\alpha_{\psi}$ and $\Delta\Phi$ have been measured. The comparison of the determined decay parameters for baryons and anti-baryons allows for three independent CP-symmetry tests: $A_{\mathrm{CP}}^{\Xi}$, $A_{\mathrm{CP}}^{\Lambda}$ and $\Phi_{\mathrm{CP}}^{\Xi}$ where the asymmetries for $\Xi$ decay are measured for the first time~\cite{abl3}.

The BESIII result for $\langle\phi_{\Xi}\rangle$ has similar precision as HyperonCP result~\cite{hua}, $\phi_{\Xi}=-0.042\pm0.016~\mathrm{rad}$, but the two values differ by 2.6 standard deviations. The $\langle\phi_{\Xi}\rangle$ measurements translates to the determination of the strong phase difference $\delta_P-\delta_S$ of $(-4.0\pm3.3\pm1.7)\times10^{-2}~\mathrm{rad}$ consistent with the heavy-baryon chiral perturbation theory predictions~\cite{tan} of $(1.9\pm4.9)\times10^{-2}~\mathrm{rad}$. The weak phase difference $(\xi_P-\xi_S)=(1.2\pm3.4\pm0.8)\times10^{-2}~\mathrm{rad}$ is in agreement with the SM predictions~\cite{tan}, $(1.8\pm1.5)\times10^{-4}~\mathrm{rad}$. This is one of the most precise tests of the CP symmetry for strange baryons and the first direct measurement of the weak phase for any baryon.  

\section{Summary and Outlook}

Hyperons provide a powerful diagnostic tool to study the strong interaction and fundamental symmetries. In particular, exclusive measurements of polarised and entangled hyperon-antihyperon pairs give access to information that is difficult or impossible to study in other processes. In recent studies by the BESIII collaboration, the structure and decay of the single-strange $\Lambda$ hyperon has been studied with unprecedented precision. The first direct measurement of the weak phase difference has been performed using the multi-strange $\Xi$ hyperon decay. Furthermore, in ongoing studies of sequentially decaying charmed hyperons, the strong and weak/beyond SM observables can be disentangled. This will, in combination with the world-record data sample of $10^{10}$ $J/\psi$ events from BESIII, has potential to bring hyperon physics to a new level.

\section*{Acknowledgments}

This work was supported in part by National Natural Science Foundation of China (NSFC) under Contract No. 11935018, the CAS President’s International Fellowship Initiative (PIFI) (Grant No. 2021PM0014) and the Polish National Science Centre through the Grant 2019/35/O/ST2/02907.

\end{document}